\documentclass[aps,prl,twocolumn,showpacs,floats]{revtex4}
\usepackage{epsfig}
\usepackage{amsmath}
\newcommand{\be}{\begin{equation}}
\newcommand{\ee}{\end{equation}}
\newcommand{\bea}{\begin{eqnarray}}
\newcommand{\eea}{\end{eqnarray}}
\newcommand{\ba}{\begin{array}}
\newcommand{\ea}{\end{array}}
\newcommand{\nn}{\nonumber}

\newcommand{\gam}{\gamma}
\newcommand{\al}{\alpha}
\newcommand{\sig}{\sigma}

\newcommand{\ua}{\uparrow}
\newcommand{\da}{\downarrow}

\newcommand{\ra}{\rangle}

\begin{document}

\title{\bf Deterministic teleportation of electrons in a quantum dot 
nanostructure}

\author{
R.L. de Visser and M. Blaauboer
}

\affiliation{Kavli Institute of Nanoscience, Delft University of Technology,
Lorentzweg 1, 2628 CJ Delft, The Netherlands
}
\date{\today}

\begin{abstract}
We present a proposal for deterministic quantum teleportation of electrons in a semiconductor nanostructure consisting 
of a single and a double quantum dot. The central issue addressed in this paper is how to design and implement the 
{\it most efficient} - in terms of the required number of single and two-qubit operations - deterministic 
teleportation protocol for this system. Using a group-theoretical analysis we show that deterministic teleportation 
requires a minimum of three single-qubit rotations and two entangling ($\sqrt{\rm swap}$) operations. 
These can be implemented for
spin qubits in quantum dots using electron spin resonance (for single-spin rotations) and exchange interaction 
(for $\sqrt{\rm swap}$ operations). 
\end{abstract}

\pacs{73.63.-b, 03.65.Ud, 03.67.Hk, 85.35.Be}
\maketitle

Quantum teleportation is the process whereby a quantum state is transported from one
place to another without moving through the intervening space. This process, proposed
by Bennett {\it et al.}~\cite{benn93}, involves a source particle 
(whose quantum state is to be transported) plus two entangled particles at a 
distant location, one of which is the target particle (to which the quantum state will be 
transferred). The teleportation procedure is then as follows: the two entangled particles 
are separated and one of them is sent to ``Alice'', who also holds the source particle, 
while the other is sent to ``Bob''. Alice performs a joint measurement 
on her two particles in the Bell basis and communicates the result to Bob. 
Depending on this information Bob applies one out of four unitary operations to his particle which 
thereby acquires the original quantum state of the source particle. If Alice can distinguish all 
four possible measurement outcomes the teleportation process can in principle be completed with 100\% success 
rate and is called deterministic. If Alice on the other hand is only able to perform a partial 
measurement on her two particles, the success probability is less than 1 and teleportation is 
probabilistic.

Probabilistic teleportation has been experimentally demonstrated for photons~\cite{bouw97} and from one 
atom to another within the same molecule~\cite{niel98}. Deterministic quantum teleportation has been achieved 
for optical fields~\cite{furu98} and atomic ions~\cite{rieb04}. Electrons have not been teleported yet. 
In recent years several theoretical proposals have appeared for quantum teleportation of electrons, 
using excitons in coupled quantum dots~\cite{rein01,chen05}, electron spins in a circuit consisting of 
normal and superconducting quantum dots~\cite{saur03}, electron-hole pairs in a two-dimensional electron 
gas~\cite{been04}, charge qubits in a quantum dot array~\cite{pasq04} and entangled electron-photon-electron 
states~\cite{leue05}. Out of these proposals, Refs.~\cite{rein01} and \cite{leue05} are in principle deterministic.

Here we propose the first all-electronic deterministic teleportation scheme for electron spins in a quantum 
dot nanostructure~\cite{others}. 
This choice of system is motivated by: (1) the high level of control over the number of electrons confined in 
quantum dots~\cite{cior00}, (2) recent advances in coherent manipulation of single and entangled pairs of spins 
in these structures~\cite{pett05} and (3) the relative robustness of the electron spin against decoherence~\cite{elze05}. 
Compared to photons and ions, however, the coherence time (T$_2$) of electron spins in quantum dots - although it
has not been measured yet for a single spin - is expected to be orders of magnitude shorter and forms the primary 
limiting factor for coherent quantum communication processes such as teleportation. The central question we therefore 
ask in this paper is how to design the {\it most efficient}, i.e. the least time-consuming, scheme for deterministic 
teleportation of electron spins. To this end, we first examine the teleportation protocol of Bennett 
{\it et al.}~\cite{benn93} and note that Alice has some freedom in the choice of her measurement basis: it does 
not need to be the Bell basis, any maximally-entangled basis will do. We then exploit this freedom to derive 
the general form of evolution operators which transform a maximally-entangled basis into the standard basis 
\{$|$$\ua \ua \ra$, $|$$\ua \da \ra$, $|$$\da \ua \ra$, $|$$\da \da \ra$\}, which is the natural measurement basis 
in quantum dots. Our key result is the demonstration that such an evolution operator (and hence Alice's full measurement)
cannot be implemented using less than one single-spin rotation plus two $\sqrt{\rm swap}$ operations. 

The paper is organized as follows. First we briefly review the original teleportation protocol~\cite{benn93} and 
propose a straightforward implementation of this protocol for probabilistic teleportation of electron spins in a 
quantum dot nanostructure. We then describe how this probabilistic scheme should be modified in order to be able 
to achieve deterministic teleportation and present our proposal. The paper ends with a discussion 
of the various time scales involved in the teleportation process and a comparison to the state-of-the-art
of relevant experimental techniques.

Starting with the scheme of Ref.~\cite{benn93}, if the source particle 1 is a spin-1/2 particle in the (unknown) 
state $| \psi\ra$ $\equiv$ $| \psi^{(1)}\ra$ = $a|$ $\ua \ra$ + $b|$ $\da\ra$ and the entangled pair (2,3) also 
consists of spin-1/2 particles in the 
singlet state $|\psi^{(2,3)}\ra$ = $|S\ra$ $\equiv$ $\frac{1}{\sqrt{2}} (|$$\ua \da\ra - |$$\da \ua\ra)$, then the 
joint state of these three particles is given by the direct product $|\psi^{(1,2,3)}\ra$ = $|\psi^{(1)}\ra 
\otimes |\psi^{(2,3)}\ra$. Rewriting $|\psi^{(1,2,3)}\ra$ in terms of the Bell basis states 
$|b_1\ra$ $\equiv$ $\frac{1}{\sqrt{2}} (|$$\ua \da\ra - |$$\da\ua\ra)$, 
$|b_2\ra$ $\equiv$ $\frac{1}{\sqrt{2}} (|$$\ua \da\ra + |$$\da\ua\ra)$, 
$|b_3\ra$ $\equiv$ $\frac{1}{\sqrt{2}} (|$$\ua \ua\ra - |$$\da\da\ra)$,
$|b_4\ra$ $\equiv$ $\frac{1}{\sqrt{2}} (|$$\ua \ua\ra + |$$\da\da\ra)$ 
for particles 1 and 2 yields $|\psi^{(1,2,3)}\ra$ = $\frac{1}{2} \left[
|b_1^{(1,2)}\ra \right. (-a|$$\ua_3\ra - b|$$\da_3\ra) +
|b_2^{(1,2)}\ra (-a|$$\ua_3\ra + b|$$\da_3\ra) $ 
$ + |b_3^{(1,2)}\ra (b|$$\ua_3\ra + a|$$\da_3\ra) +
|b_4^{(1,2)}\ra (-b|$$\ua_3\ra + a|$$\da_3\ra) \left. \right]$. If Alice measures 
particles 1 and 2 in the Bell basis, Bob can reconstruct
the original state $|\psi \ra$ by applying an appropriate unitary transformation 
to his particle 3.

\begin{figure}
\centerline{\epsfig{figure=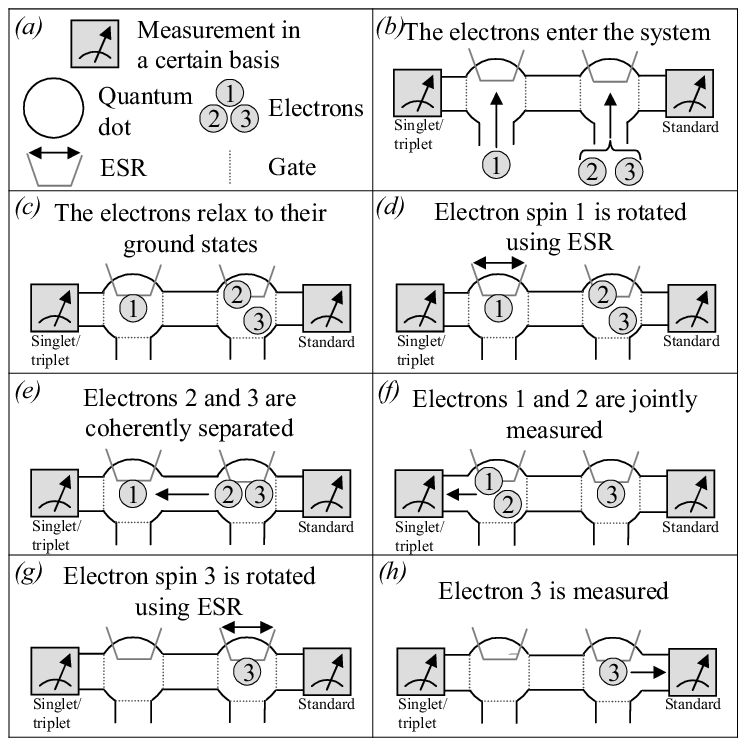,height=8.cm,width=0.9\hsize}}
\caption[]{Schematic set-up for probabilistic quantum teleportation of an electron spin. 
(a): Explanation of symbols used, ESR= electron spin resonance. (b)-(d): Initialization. 
(e)-(f): Joint measurement of spins 1 and 2 in the singlet-triplet basis. 
(g): Reconstruction. (h): Read-out.} 
\label{fig:prob}
\end{figure}

{\it Probabilistic teleportation}. -- Fig.~\ref{fig:prob} shows an implementation of 
this protocol for probabilistic teleportation of electron spins in a nanostructure 
consisting of two quantum dots. The electron spins are prepared in 
their respective ground states $|\psi^{(1)}\ra = |$$\ua\ra$ and $|$$\psi^{(2,3)}\ra = |S\ra$~\cite{hans04} 
[Fig.~\ref{fig:prob}(c)]. In order to convincingly demonstrate successful teleportation, 
one should be able to initialize electron 1 in any arbitrary superposition of spin states 
(not only the ground state $|$$\ua\ra$). This can be achieved by coherently rotating 
the spin of electron 1 using ESR [Fig.~\ref{fig:prob}(d)]
which results in $|\psi^{(1)}\ra = a |$$\ua\ra + b|$$\da\ra$. In the measurement depicted in
Fig.~\ref{fig:prob}(f) Alice can 
only distinguish the singlet from the triplets. If her measurement outcome is a singlet,
which occurs with probability 1/4, the teleportation process can be completed successfully. 
In order to verify that teleportation has indeed been successful, Bob can 
apply the inverse coherent rotation of the one used for initialization in Fig.~\ref{fig:prob}(d) 
to his spin which should then be measured with certainty in the $|$$\ua\ra$ state 
[Fig.~\ref{fig:prob}(h)]. Note that this scheme is a direct electronic 
analogue of the first demonstration of teleportation with polarization-entangled
photons by Bouwmeester {\it et al.}~\cite{bouw97}.

{\it Deterministic teleportation}. --  In order to achieve deterministic quantum 
teleportation, Alice has to perform a full Bell measurement on her two particles. 
For electron spins
in quantum dots, however, no measurement technique is available for full Bell
measurements: current techniques allow to distinguish singlet from triplets and to perform 
full measurements in the standard basis~\cite{elze04,hans05}. We therefore use an idea 
proposed by Brassard {\it et al.}~\cite{bras98}, which consists of first transforming 
from the Bell basis to the standard basis and then measuring the spins in the 
latter basis~\cite{approach}. The main part of the following analysis is devoted to finding 
the most efficient implementation of this procedure. We first note that the 
original protocol Ref.~\cite{benn93} is a special case of a more general teleportation 
protocol. This general protocol allows Alice to freely choose her measurement basis, 
as long as it 
is a maximally-entangled basis. The reason behind this freedom is that any 
maximally-entangled basis can be used to measure the particles 1 and 2 without gaining any 
information on the state of the source particle. This freedom can be exploited to obtain a 
more efficient sequence of operations for Alice's measurement. We now first describe the 
general form of a two-qubit operator $U \in SU (4)$ which transforms the standard basis into 
a maximally-entangled basis (in our teleportation scheme we need $U^{\dag}$). We then 
derive the most efficient translation of this general form into single-qubit rotations and 
two-qubit ($\sqrt{\rm swap}$) operations~\cite{burk99}.

Any operator $U \in SU (4)$ which transforms the standard basis into 
a maximally-entangled basis has one of the forms $U_1$, $U_2$, or $U_3$, where
\begin{subequations}
\bea
U_1 & \equiv & K_1 e^{\frac{1}{4}i\pi \sig_x \otimes \sig_x} R_x^{(1)}(\mu) R_x^{(2)}(\nu)
R_z^{(1)}(\xi) R_z^{(2)}(\zeta) \label{eq:Ux} \\
U_2 & \equiv & K_1 e^{\frac{1}{4}i\pi (\sig_x \otimes \sig_x + \epsilon \sig_z \otimes \sig_z)} 
R_z^{(1)}(\xi) R_z^{(2)}(\zeta) \ \ \epsilon\! \in\! (0,1) \label{eq:Uy} \\
U_3 & \equiv & K_1 e^{\frac{1}{4}i\pi (\sig_x \otimes \sig_x + \sig_z \otimes \sig_z)} 
R_y^{(1)}(\mu) R_y^{(2)}(\nu) R_z^{(1)}(\xi) R_z^{(2)}(\zeta)
\label{eq:Uz}
\eea
\label{eq:U}
\end{subequations}
with $K_1$ $\in$ $SU(2)$ $\otimes$ $SU(2)$, $\sig_{i}$ $(i=x,y,z)$ the Pauli matrices, $R_n^{(1)}(\mu)$ 
a rotation of qubit 1 by an angle $\mu$ around axis 
$n$ and $\mu, \nu, \xi,\zeta \in [0,2\pi)$. The mathematical proof of Eq.~(\ref{eq:U})
makes use of the theory of Lie groups and Lie algebra's and relies on the so-called 
K$_1$AK$_2$ decomposition theorem, which says that every $U\in SU(4)$ can be decomposed as $K_1 
e^{i \al \sig_x \otimes \sig_x + i \beta \sig_y \otimes \sig_y + i \gam \sig_z \otimes \sig_z} K_2$,
with $K_1, K_2 \in SU(2) \otimes SU(2)$ and $\al, \beta, \gamma \in [0,2\pi)$~\cite{khan01}.
The full proof of Eq.~(\ref{eq:U}) is given in Ref.~\cite{viss06}. Here we proceed to translate 
the operators in Eq.~(\ref{eq:U}) into the shortest possible sequence of single-spin rotations and 
$\sqrt{\rm swap}$ operations. Starting with the former, $R_n^{(1)}(\al)$ and $R_n^{(2)}(\al)$
are given by
\be
R_n^{(1)}(\al) = e^{-\frac{1}{2}i\al \vec{n}\cdot \vec{\sig} \otimes I }\ , \ 
R_n^{(2)}(\al) = e^{-\frac{1}{2}i\al I \otimes \vec{n}\cdot \vec{\sig} }
\label{eq:rotations}
\ee
with $\vec{n}$ $\equiv$ $(\sin \theta \cos \phi,\sin \theta \sin \phi,\cos\theta)$
a unit vector on the Bloch sphere [$\theta$$\in$$[0,\pi)$, $\phi$$\in$$[0,2\pi)$], 
$\vec{\sig}\equiv (\sig_x,\sig_y,\sig_z)$ and $I$ the identity matrix. 
Eq.~(\ref{eq:rotations}) originates from the evolution operator $U_R(t)=
\exp [-(i/\hbar)\int_0^t H_R(\tau) d\tau]$ and corresponds to an electron-spin-resonance (ESR)
pulse applied to a single spin which is described by
the Hamiltonian ${\mathcal H}_R(t) = - \frac{1}{2}\hbar \gam \vec{B}(t)\cdot \vec{\sig}$ 
[$\gam$ is the gyromagnetic ratio and $\alpha \equiv -\gamma \int_0^{t} B(\tau) d\tau$]. 
Similarly, the $\sqrt{\rm swap}$-operation for electron spins in quantum dots originates 
from the exchange interaction described by the Heisenberg Hamiltonian
${\mathcal H}_{EX}(t) = (1/4) \hbar^2 J(t)\, \vec{\sig}^{(1)} \cdot \vec{\sig}^{(2)}$,
with $J(t)$ the time-dependent exchange energy, the energy difference between singlet and 
triplets. The time evolution operator $U_{EX}(t)$ corresponding to
${\mathcal H}_{EX}(t)$, or equivalently $EX(\beta) \equiv U_{EX}(t)$, is given by
\be
EX(\beta) = e^{-\frac{1}{4}i\beta} \left(
\ba{cccc}
e^{\frac{1}{2}i\beta} & 0 & 0 & 0 \\
0 & \cos (\frac{1}{2}\beta) & i\sin (\frac{1}{2}\beta) & 0 \\
0 & i \sin (\frac{1}{2}\beta) & \cos (\frac{1}{2}\beta) & 0 \\
0 & 0 & 0 & e^{\frac{1}{2}i\beta}
\ea \right). \nn
\ee
Here $\beta(t) \equiv -\hbar \int_0^t J(\tau) d\tau$. $EX(\pi)$ represents
the swap-operation which exchanges the states of two spins. An interaction switched on
for half this time, $EX(\pi/2)$, is called the $\sqrt{\rm swap}$ operation. 

\begin{figure}
\centerline{\epsfig{figure=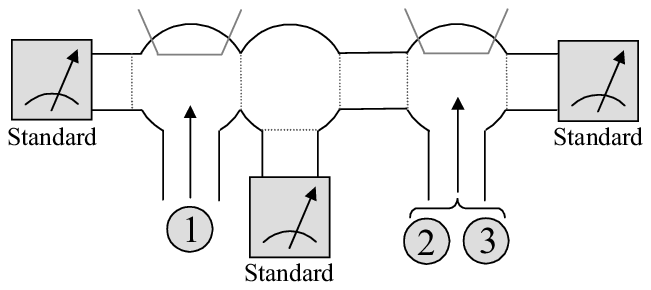,height=2.8cm,width=0.79\hsize}}
\caption{Quantum dot set-up for deterministic teleportation of an electron. 
See the text for explanation.} 
\label{fig:det}
\end{figure}

Regarding $U_1$-$U_3$ in Eq.~(\ref{eq:U}), we first note that $\exp[(1/4)i\pi \sig_x \otimes \sig_x]$ 
cannot be expressed 
as a single $\sqrt{\rm swap}$ operation plus rotations. This can directly be seen from the 
form of the matrix $(\sqrt{\rm swap})^{\dag} \exp[(1/4)i\pi \sig_x \otimes \sig_x]$.
$U_1$-$U_3$ can, however, be expressed in terms of two $\sqrt{\rm swap}$ operations as
\be
U \equiv \sqrt{\rm swap}\, R_{n_{\phi}}^{(1)}(\pi)\, \sqrt{\rm swap}
\label{eq:mainresult}
\ee
with $\vec{n}_{\phi} \equiv (\cos \phi, \sin \phi, 0)$~\cite{choice}. 
Eq.~(\ref{eq:mainresult}) is the {\it optimal}
decomposition of Eq.~(\ref{eq:U}) in terms of the number of required operations~\cite{zhan03}.  
Rewriting $U$ as
$U^{'} \equiv - \exp[-(1/4)i\pi]\,U$, we find that $U^{'}$ transforms the standard
basis into the maximally-entangled basis
\begin{subequations}
\bea
U^{'} |\! \ua \ua \ra & = & \frac{1}{\sqrt{2}} e^{i\phi}\, (i|\! \ua \da\ra + |\! \da \ua\ra) \\
U^{'} |\! \ua \da \ra & = & \frac{1}{\sqrt{2}} (i e^{-i\phi}|\! \ua \ua\ra + e^{i\phi} |\! \da \da\ra) \\
U^{'} |\! \da \ua \ra & = & \frac{1}{\sqrt{2}} (e^{-i\phi}|\! \ua \ua\ra + i e^{i\phi}|\! \da \da\ra) \\
U^{'} |\! \da \da \ra & = & \frac{1}{\sqrt{2}} e^{-i\phi}\, (|\! \ua \da\ra + i |\! \da \ua\ra).
\eea
\end{subequations}
After Alice has applied $U^{'\dagger}$ to her two qubits, the appropriate unitary transformations 
which Bob has to perform in order to reconstruct the original quantum state can be calculated using
the approach in Ref.~\cite{liscriptie} and are given by ($Q_1$ corresponds to measurement 
outcome $|$$\ua \ua\ra$, $Q_2$ to $|$$\ua \da\ra$ etc.)
\begin{subequations}
\bea
& Q_1 = \left( \ba{cc}
-i & 0 \\
0 & 1 \ea \right)  &
Q_2 = \left( \ba{cc}
0 & i e^{-i\phi} \\
-e^{i\phi} & 0 \ea \right) \nn \\
Q_3 & = \left( \ba{cc}
0 & e^{-i\phi} \\
-i e^{i\phi} & 0 \ea \right)  &
Q_4 = \left( \ba{cc}
-1 & 0 \\
0 & i \ea \right). \nn
\eea
\end{subequations} 
Translating $Q_i$ into single-spin rotations in the ($x,y$)-plane~\cite{xyplane} we obtain 
$Q_1=R_{n_{\phi}}(\pi/2)R_{n_{\phi}^{\perp}}(\pi/2)R_{n_{\phi}}(-\pi/2)$ and similar results for $Q_2-Q_4$, see 
Table~\ref{table:Bobrotations}.
\begin{table}
\begin{tabular}{|c|c|c|c|}
\hline \hline
Alice's & Reconstruction of & Measure & Measure \\ 
result & original state & $|$$\ua\ra$ & $|$$\da\ra$ \\ \hline
$|$$\ua \ua \ra$ & $R_{n_{\phi}}(\frac{\pi}{2}) R_{n_{\phi}^{\perp}}(\frac{\pi}{2}) R_{n_{\phi}}(-\frac{\pi}{2})$ & 
$R_{n_{\phi}^{\perp}}(\rho)$ & $R_{n_{\phi}^{\perp}}(\rho + \pi)$ \\ \hline
$|$$\ua \da \ra$ & $R_{n_{\phi}}(\frac{\pi}{2}) R_{n_{\phi}^{\perp}}(\frac{\pi}{2}) R_{n_{\phi}}(\frac{\pi}{2})$ & 
$R_{n_{\phi}^{\perp}}(-\rho + \pi)$ & $R_{n_{\phi}^{\perp}}(-\rho)$ \\ \hline
$|$$\da \ua \ra$ & $R_{n_{\phi}}(\frac{\pi}{2}) R_{n_{\phi}^{\perp}}(-\frac{\pi}{2}) R_{n_{\phi}}(\frac{\pi}{2})$ & 
$R_{n_{\phi}^{\perp}}(\rho + \pi)$ & $R_{n_{\phi}^{\perp}}(\rho)$\\ \hline
$|$$\da \da \ra$ & $R_{n_{\phi}}(-\frac{\pi}{2}) R_{n_{\phi}^{\perp}}(\frac{\pi}{2}) R_{n_{\phi}}(\frac{\pi}{2})$ & 
$R_{n_{\phi}^{\perp}}(-\rho)$ & $R_{n_{\phi}^{\perp}}(-\rho + \pi)$\\ \hline \hline
\end{tabular}
\caption{Spin rotations in the (x,y)-plane which Bob has to apply as a function of Alice's measurement outcome
(first column) in order to reconstruct the original quantum state (second column), measure with certainty $|$$\ua\ra$ 
(third column) or $|$$\da\ra$ (fourth column) if particle 1 has been rotated by an angle $\rho$ 
[Fig.~(\ref{fig:prob}(d)]. $n_{\phi}^{\perp}$ denotes the axis in the (x,y)-plane perpendicular to $n_{\phi}$.
}
\label{table:Bobrotations}
\end{table}
By applying these transformations, the original quantum state is reconstructed and remains available 
for further processing. In order to demonstrate that the teleportation process has been successful,
one could perform an analogous experiment as depicted in Fig.~\ref{fig:prob}(g), where Bob rotates the spin 
of his particle after receiving Alice's message such that it 
ends up as $|$$\ua\ra$ or $|$$\da\ra$. He should then 
measure with certainty $|$$\ua\ra$ resp. $|$$\da\ra$. The appropriate rotations Bob needs to apply in this case 
are also given in Table~\ref{table:Bobrotations}. The quantum state of the source particle is destroyed 
in this experiment, so it cannot be used for further processing. Fig.~\ref{fig:det} shows 
how the set-up of Fig.~\ref{fig:prob} has to be modified for deterministic teleportation. All steps of 
the protocol are the same as in Fig.~\ref{fig:prob}, except for Alice's measurement in Fig.~\ref{fig:prob}(f), 
which now consists of the 
operation $U$ [Eq.~(\ref{eq:mainresult})] followed by read-out of the two spins in the standard basis. 
Note that this scheme is in principle also suitable for
 entanglement swapping~\cite{yurk92} if one starts out with two pairs of entangled spins in 
separate dots: by bringing together and entangling one spin of each pair and performing a 
Bell measurement on this newly formed pair the remaining (separated) spins, which have never met, 
become entangled.

{\it Feasibility}.-- In this last section we discuss the feasibility of the proposed teleportation schemes. We
first note that all of the ingredients (double quantum dots occupied by two entangled electron 
spins~\cite{cior00}, ESR rotations~\cite{kopp06}, $\sqrt{\rm swap}$-operations~\cite{pett05}, single-shot read-out 
techniques for electron 
spins~\cite{elze04,hans05,detector}) are already available. 
Comparing the duration of
an ESR rotation $\sim 50$ ns for magnetic fields of 1 mT~\cite{vand04} to the duration of a 
$\sqrt{\rm swap}$ operation $\sim$ 180 ps~\cite{pett05} and assuming fast 
read-out we see that the total time duration of the teleportation 
process is mainly determined by the ESR rotations. For the probabilistic scheme 
we then find $t_{\rm prob} \sim$ 100 ns (corresponding to 
two rotations) and for the deterministic scheme $t_{\rm det} \sim$ 150-250 ns 
(depending on whether Bob measures the spin of his particle or not).
Teleportation can thus be completed within the single-spin relaxation time
$T_1 \sim 1$ ms~\cite{elze04}. Whether it can also be completed before decoherence 
sets in is as yet unknown. Recent measurements indicate that 
$T_2$ is limited by hyperfine interaction in a GaAs quantum dot and report times
$T_2^{*} \sim 10$ ns~\cite{pett05,kopp05}. However, using quantum control techniques such as spin echo,
these times can be extended to $> 1\, \mu$s~\cite{pett05}. This technique has also been
employed in one of the teleportation experiments with ions~\cite{rieb04}.

In conclusion, we have proposed a scheme for deterministic teleportation of
electrons in a quantum dot nanostructure which involves the {\it shortest possible}
sequence of single-spin rotations and two-qubit ($\sqrt{\rm swap}$) operations. 
Due to the high level of control over the system, teleportation can
in principle be achieved with 100\% success rate at the "push of a button", similar as in 
the experiments with ions~\cite{kimb04}. To actually perform the proposed experiment
is challenging, mainly due to the required fast coherent rotations and read-out, but does not
seem unfeasible.

Stimulating discussions with L.P. Kouwenhoven, L.M.K. Vandersypen and I.T. Vink
are gratefully acknowledged. This work has been supported by the Netherlands Organisation for
Scientific Research (NWO) and by the EU's Human Potential Research Network under
contract No. HPRN-CT-2002-00309 (``QUACS'').


\begin{thebibliography}{99}
\bibitem{others}
but the scheme can also be applied to other systems with three qubits.
\bibitem{approach}
This approach was also used in the NMR and ion experiments, 
Refs.~\cite{niel98} and \cite{rieb04}.
\bibitem{zhan03}
See also J. Zhang {\it et al.}, Phys. Rev. Lett. {\bf 91}, 027903 (2003), in which an
upper bound is derived for the number of two-qubit operations that are required 
for implementing quantum gates of the form $\exp [i\gamma \sigma_z \otimes \sigma_z]$ 
with $\gamma \in (0,\pi/4]$.
\bibitem{xyplane}
Here we assume the Zeeman-splitting field to be in the $z$-direction, so that 
spin rotations are performed in the (x,y)-plane.
\bibitem{detector}
In the probabilistic scheme, Alice's measurement can be done straightforwardly 
by tuning the energy levels in the left dot in Fig.~\ref{fig:prob}(e) such that only the singlet level 
is available for the incoming electron 2.
\end{thebibliography}
\end{document}